\documentclass[preprint,superscriptaddress,twocolumn,lengthcheck,tightenlines,showpacs,a4paper,10pt,prl]{revtex4}
\usepackage{amssymb}
\usepackage{amsfonts}

\input{tcilatex}
\begin{document}

\preprint{APS/123-QED}
\title{ Spin Liquid State in the 3D Frustrated Antiferromagnet PbCuTe$_{2}$O$%
_{6}$: NMR and $\mu $SR Studies}
\author{P. Khuntia*}
\affiliation{Ames Laboratory, US Department of Energy, Ames, Iowa 50011, USA}
\affiliation{Laboratoire de Physique des Solides, CNRS, Univ. Paris-Sud, Universit\'{e}
Paris-Saclay, 91405 Orsay Cedex, France}
\author{F. Bert}
\affiliation{Laboratoire de Physique des Solides, CNRS, Univ. Paris-Sud, Universit\'{e}
Paris-Saclay, 91405 Orsay Cedex, France}
\author{P. Mendels}
\affiliation{Laboratoire de Physique des Solides, CNRS, Univ. Paris-Sud, Universit\'{e}
Paris-Saclay, 91405 Orsay Cedex, France}
\author{B. Koteswararao}
\affiliation{School of Physics, University of Hyderabad, Central University PO, Hyderabad
500046, India}
\affiliation{Center of Condensed Matter Sciences, National Taiwan University, Taipei
10617, Taiwan}
\author{A. V. Mahajan}
\affiliation{Department of Physics, Indian Institute of Technology Bombay Powai
Mumbai-400076, India}
\author{M. Baenitz}
\affiliation{Max Planck Institute for Chemical Physics of Solids, 01187 Dresden, Germany}
\author{F. C. Chou}
\affiliation{Center of Condensed Matter Sciences, National Taiwan University, Taipei
10617, Taiwan}
\author{C. Baines}
\affiliation{Laboratory for Muon Spin Spectroscopy, Paul Scherrer Institute, CH-5232
Villigen PSI, Switzerland}
\author{A. Amato}
\affiliation{Laboratory for Muon Spin Spectroscopy, Paul Scherrer Institute, CH-5232
Villigen PSI, Switzerland}
\author{Y. Furukawa}
\affiliation{Ames Laboratory, US Department of Energy, Ames, Iowa 50011, USA}
\affiliation{Department of Physics and Astronomy, Iowa State University, Ames, Iowa
50011, USA}

\begin{abstract}
PbCuTe$_{2}$O$_{6}$ is a rare example of a spin liquid candidate featuring a
three dimensional magnetic lattice. Strong geometric frustration arises from
the dominant antiferromagnetic interaction which generates a hyperkagome
network of Cu$^{2+}$ ions although additional interactions enhance the
magnetic lattice connectivity. Through a combination of magnetization
measurements and local probe investigation by NMR and $\mu $SR down to 20
mK, we provide a robust evidence for the absence of magnetic freezing in the
ground state. The local spin susceptibility probed by the NMR shift hardly
deviates from the macroscopic one down to 1~K pointing to a homogeneous
magnetic system with a low defect concentration. The saturation of the NMR
shift and the sublinear power law temperature (\textit{T})\ evolution of the
1/$T_{1}$ NMR relaxation rate at low \textit{T} point to a non-singlet
ground state favoring a gapless fermionic description of the magnetic
excitations. Below 1~K a pronounced slowing down of the spin dynamics is
witnessed, which may signal a reconstruction of spinon Fermi surface.
Nonetheless, the compound remains in a fluctuating spin liquid state down to
the lowest temperature of the present investigation.
\end{abstract}

\volumeyear{year}
\volumenumber{number}
\issuenumber{number}
\eid{identifier}
\date[Date text]{: updated version on: }
\received[Received text]{date}
\revised[Revised text]{date}
\accepted[Accepted text]{date}
\published[Published text]{date}
\startpage{1}
\endpage{2}
\maketitle

Combining competing interactions and quantum fluctuations, maximized for low
spin \textit{S} = 1/2, is one major track followed in the past decade to
discover novel disordered quantum states beyond the Landau paradigm of phase
transition with broken symmetries. One such long sought state is a quantum
spin liquid (QSL), breaking no symmetries down to $T=0$ but exhibiting
macroscopic entanglement of strongly interacting spins and featuring exotic
fractionalized excitations~\cite{LB,CL}. Much effort has been devoted
towards low dimensional quantum antiferromagnets (AFM) where low lattice
coordination helps in further destabilizing classical ground states in favor
of more exotic ones driven by quantum fluctuations~\cite{SS}. A few spin
liquid materials have been identified, either based on the highly frustrated
kagome lattice - a weakly coordinated ($z=4)$ network made of corner-sharing
triangles - including the celebrated Herbertsmithite mineral~\cite{PM,AO,TH}%
, or based on the less frustrated simple triangular lattices ($z=6$)~\cite%
{YSZ,RK}. In three dimensional (3D) systems quantum states are even more
elusive. In the double perovskite Ba$_{2}$YMoO$_{6}$ compound featuring
edge-shared tetrahedra ($z=12$), the Mo$^{5+}$ ions host nearly pure spin $%
S=1/2$ which quench partially into a valence bond glass state \cite{TA,MDV}.
Further developments in this context arise from some rare-earth based
pyrochlores where large spins with strong anisotropies decorate
corner-sharing tetrahedra ($z=6$). The low energy physics can be mapped to a
model of interacting effective spin \textit{S} = 1/2 and may stabilize QSL
states~\cite{gingras}. The hyperkagome structure - a 3D network of
corner-sharing triangles ($z=4$) offers an interesting alternative and
indeed several exotic quantum phases relying on this geometry have been
proposed~\cite{MJLa}. At the origin of these studies is the Na$_{4}$Ir$_{3}$O%
$_{8}$ compound \cite{HT} where Ir$^{4+}$ ions bear effective $J_{\mathrm{eff%
}}$ $=1/2$ spins and fail to order well below the exchange interaction
energy although the ground state has recently been shown to be static \cite%
{MJG,ACS}. Whether perturbation terms to the Heisenberg model, such as
exchange anisotropies, or disorder in the interaction is responsible for the
low $T$ freezing is still an open issue. Remarkably, those spin liquid
candidates in 3D are based on rather heavy ions where the applicability of a
model of \textit{S} = 1/2 Heisenberg spins is questionable.

Recently, some of us reported a quantum AFM PbCuTe$_{2}$O$_{6}$ (henceforth
PCTO), which constitutes a strongly frustrated 3D network of Cu$^{2+}$ ($S$
= 1/2 ) spins. PbCuTe$_{2}$O$_{6}$ crystallizes in a cubic structure with a
hyperkagome magnetic lattice if only the dominant second nearest neighbor ($%
n.n.$) interaction $J$ is considered~\cite{3D}. Additionally, weaker 1st $%
n.n.$ ($\sim 0.5J$) and 3rd $n.n.$ ($\sim 0.8J$) AFM interactions form
isolated triangles and chains. The magnetic susceptibility $\chi $\ exhibits
\ a Curie-Weiss (CW) behavior with an AFM $\theta _{CW}\simeq -22K$ and no
sign of magnetic transition down to 2~K. The magnetic specific heat ($C_{m}$%
) shows a broad maximum at $T^{max}\simeq 0.05\theta _{CW}\simeq 1.1$~K,
quite similarly as in Na$_{4}$Ir$_{3}$O$_{8}$, followed by a weak kink at
0.87~K of unclear origin. In magnetic fields larger than 8~T, the evolution
of $C_{m}$ with temperature is nearly quadratic at low \textit{T}, in line
with some theoretical predictions for the quantum hyperkagome model~\cite%
{MJLb,YZhou}.

In this Letter, we present a comprehensive account of the local magnetic
susceptibility and low temperature spin dynamics via NMR and $\mu $SR
measurements accompanied by low temperature magnetization studies on the
highly frustrated 3D quantum antiferromagnet PbCuTe$_{2}$O$_{6}$. $\mu $SR
data reveal no signature of long range magnetic ordering (LRO) down to
20~mK, a hallmark of a QSL state. The persistence of slow spin dynamics is
confirmed by the NMR signal intensity being wiped out below 1~K and not
recovered down to our lowest temperature \textit{T} = 50~mK. Before the
signal is lost, the NMR shift saturates at a finite value pointing to a
non-singlet ground state.\ \ \ \ \ \ \ \ \ \ \ \ \ \ \ \ \ \ \ \ \ \ \ \ \ 

Polycrystalline PCTO sample was synthesized by the method described in Ref. 
\cite{3D}. Fig. 1 shows the \textit{T}-dependence of magnetic
susceptibility. The magnetic susceptibility ($\chi =M/H$) displays a CW
behavior at high-$T$ and an enhancement at low temperature without any clear
signature of LRO. A weak ZFC-FC splitting is nonetheless observed at 0.87 K
at the lowest applied field of 0.01~T. At higher magnetic fields the
irreversibility is suppressed but a kink remains detectable which slightly
shifts towards lower \textit{T}. This magnetic anomaly has to be connected
to the kink in $C_{m}$ at the same temperature (see inset of Fig.1). The
relevance of this spin-glass like transition for the bulk properties of PCTO
is difficult to decide on the sole basis of the macroscopic measurements. In
the following we use local probe measurements and clearly demonstrate that
the 0.87 K anomaly is not a bulk transition but is attributed to the
presence of a tiny amount of defects in the polycrystaline sample.

To gain further insights into the spin dynamics and ground state properties,
we have performed $\mu $SR measurements at Paul Scherrer Institute. The zero
field relaxation of the polarization of the muons stopped in the sample
could be fitted to a single stretched exponential model \textit{P}$_{z}$(%
\textit{t}) = exp[-($\lambda $\textit{t})$^{\beta }]$ in the whole
temperature range (see Supplementary Material \cite{SM}). The monotonic
decay of the polarization even at the lowest $T$= 20 mK demonstrates the
absence of static internal field. In particular, the characteristic
signatures of a frozen ground state, namely (damped) spontaneous
oscillations and for a powder sample a non-zero polarization at long time
due to internal fields directed along the initial muon polarization, are not
observed in PCTO. The transition to static magnetism seen at 0.87~K in bulk
magnetization measurements should therefore be attributed to a minority spin
fraction undetected in the $\mu $SR experiment, $i.e.$ below a few percent
of the sample volume. As detailed below, the fact that bulk spins slow down
on the verge of static magnetism in this same temperature range, suggests
nevetheless that the minority spin fraction does not constitute a separated
impurity phase but could arise from some slightly disordered areas/grains in
the polycrystalline sample.

The evolution of the relaxation rate ($\lambda $) and the stretched exponent 
$\beta $ are shown in figure 2(b). The relaxation is close to exponential
and hardly dependent on temperature above about 5 K indicating that the
system is close to its paramagnetic limit \cite{TM}. At lower temperature,
the increase of the relaxation rate renders evidence for a slowing down of
the spin dynamics likely resulting from the building up of short range
correlations. Upon further cooling below about 1 K, the increase steeply
accelerates as if on the verge of a magnetic transition but then levels off
below about 0.6 K. Such a saturation of $\lambda $ is a common feature of
highly frustrated magnets signaling the persistence of slow spin dynamics at 
$T\rightarrow 0$ in line with a QSL ground state.

The evolution of the relaxation shape from exponential ($\beta \sim $ 1) to
Gaussian ($\beta \sim $ 2) across $\sim 1$~K suggests that the electron spin
fluctuations have slowed down substantially in the ground state at the limit
of static magnetism. To quantify the level of fluctuations in the ground
state, we have fitted the relaxation (see Fig. 2(a)) to the dynamical
Kubo-Toyabe model \cite{Hayano79} $P_{DKT}(t,\Delta H,\nu ,H_{LF})$ which
accounts for a gaussian distribution of internal fields of width $\Delta H$
fluctuating at the rate $\nu $, in zero field or with an applied
longitudinal field $H_{LF}$. In zero field, this model accounts well for the
relaxation and gives $\Delta H=1.1~$mT and $\nu =0.7$~MHz. With a ratio $%
r=\gamma _{\mu }\Delta H/\nu \sim 1.3$ (where $\gamma _{\mu }=2\pi \times
135.5$~Mrad/s is the muon gyromagnetic ratio), the Cu$^{2+}$ spin
fluctuations seem indeed to have slowed down to the quasi-static limit ($%
r\sim 1$) at base temperature. However, keeping these zero-field parameters,
the model fails to account for the field dependence. Indeed in case of
(quasi-)static magnetism one expects a strong reduction of the relaxation
under an applied field of $\sim 5\Delta H$\ \ $\approx $ 5 mT.
Experimentally a field at least 20 times larger is needed to reach a similar
reduction, implying a more dynamical scenario (Fig. 2(a)). Also surprising
is the magnitude of the internal field, $\sim 1$~mT which corresponds to a
tiny moment $\sim 0.065\mu _{B}$ per Cu$^{2+}$ ions obtained by estimating
the dipolar field at the muon assumed to stop close to an oxygen site. These
two features are strongly reminiscent of the "sporadic" model introduced to
explain the "undecouplable gaussian shape" observed in the kagome bilayers
chromates~\cite{YJM,DB}. This model assumes that the relaxation in the spin
liquid state arises mostly from deconfined spinon excitations which pass
close to the muon for only a fraction of time $ft$ while the background
ground state is hardly magnetic, if not a singlet state, giving no sizeable
relaxation for the remaining time fraction $(1-f)t$. This results in a
renormalization of the parameters of the dynamical Kubo Toyabe \textit{P}$%
_{z}$(\textit{t}) = \textit{P}$_{DKT}$(\textit{t}, $f\Delta H,fH_{LF},f\nu )$%
. From the field dependence of \textit{P}$_{z}$(\textit{t}), we estimate $%
f\sim 1/10$. A detailed comparison calls for specific measurements versus
field and temperature which is beyond the scope of the present study, but
gives a direction for further $\mu $SR studies.

In complex systems such as frustrated magnets where static magnetism and
persistent fluctuations are often found to coexist at low temperatures, the
comparison of different techniques with different time windows is necessary
to get a comprehensive understanding of the ground state properties. In
addition to $\mu $SR experiments, $^{207}$Pb (\textit{I} = 1/2; \ $\gamma
_{n}/2\pi $=8.874 MHz/T) NMR measurements were carried out. Shown in Fig.
3(a) are the field swept $^{207}$Pb NMR\ spectra of \ PCTO at 63.5 MHz at
different temperatures. The absence of major structural distortions/defects
and a single $^{207}$Pb nuclear site in the host lattice result in narrow
spectra. This offers the opportunity to track the local magnetic
susceptibility unambiguously. Shown in Fig. 3(b) is the integrated NMR
signal intensity, after taking into account the spin-spin relaxation \textit{%
T}$_{2}$ correction \cite{SM}. Remarkably, the intensity decreases
drastically below 1.6 K suggesting very fast relaxation times of $^{207}$Pb
nuclei on the time scale of the NMR window, which is attributed to the
slowing down of Cu$^{2+}$ spins at low \textit{T}. Such a wipe out of the
NMR signal has been observed in a few cases and ususally the signal is
recovered at low \textit{T}, below a peak of $1/T_{1}$ at the transition
temperature resulting from the critical slowing down of the spin dynamics 
\cite{TI,RW,LL,FB}. At variance, here we did not recover the NMR\ signal
intensity even at 50 mK implying the persistence of slow spin dynamics at
very low \textit{T}. This is in perfect agreement with the $\mu $SR data and
confirms the dynamical nature of the ground state. The tiny broad signal
detected at 0.7~K may then be related to the minority spin fraction
undergoing the magnetic transition at 0.87~K. The gaussian line shape of
this remaining signal suggests a disordered, spin-glass like state for these
spins in line with the observed hysteresis of the magnetization -- as one
would expect a rectangular shaped powder average spectrum in a LRO phase~%
\cite{AFM}. From now on, we will concentrate on the NMR results in the $T$%
-range $T>1$~K where all of the bulk spins are probed.

At high temperature, the NMR line shift $^{207}$\textit{K} scales with the
macroscopic susceptibility $\chi $ : $^{207}K$ =$\frac{A_{hf}}{N_{A}}\chi
+K_{0}$, where $A_{hf}$ is the hyperfine coupling constant and represents
the hyperfine interaction between Cu electron spin and $^{207}$Pb, \textit{N}%
$_{A}$ is the Avogadro's number, and $K_{0}$ is the \textit{T-}independent
chemical shift. We obtain $A_{hf}$ = (1$\pm $ 0.05) T/$\mu _{B}$ and $K_{0}$%
= --0.05\% from a linear fit of $^{207}K$ vs $\chi $. The \textit{T}%
-dependent part of the NMR line shift, \textit{K}$_{spin}$, proportional to
the spin part of the local susceptibility, is shown in Fig.4(a). The $\chi $
and $K_{spin}$ are well reproduced by the high temperature series expansion
(HTSE)\ and (7,7), (8,7) Pad\'{e} approximants for the Heisenberg model on
the hyperkagome lattice with an AFM coupling strength of $J$/$k_{B}$ $%
\approx $ (14 $\pm $ 1)K between Cu spins~\cite{SM, HTSE}.\ Below about
10~K, the local susceptibility tracked by $^{207}$\textit{K} slightly
deviates from the macroscopic one, pointing to the contribution of a tiny
0.4(3)\% fraction of quasi-free spin (or so called orphan spins) to the
latter, and is almost constant down to 1~K \cite{FN}. The saturation at a
finite and rather a high value of the local susceptibility at low \textit{T}
is quite similar to the case of Na$_{4}$Ir$_{3}$O$_{8}$ and contrasts with
many 2D frustrated AFM where the local susceptibility exhibits a broad
maximum at a fraction of \textit{J}/\textit{k}$_{B}$~\cite{AO,JAQ1}.
Furthermore, given that no large deviation is observed between the
macroscopic $\chi $ and $^{207}$\textit{K} down to 1~K one can infer from
the magnetization data (shown in Fig.~1) that the intrinsic susceptibility
does not vary much either below 1~K. In particular, it seems unlikely that a
spin gap larger than $\sim 0.45$~K~($\approx \theta /50)$ opens up which is
confirmed by the existence of fluctuating local fields at temperatures as
low as 50~mK.

Further insight into the spin correlations is provided by the $^{207}$Pb
spin-lattice relaxation $T_{1}$ measurements. As shown in Fig. 4(b) 1/$T_{1}$
varies rather weakly with temperature, decreasing by a factor $\sim 3$ from
its maximum at 300~K down to its minimum at 2~K, ruling out the possibility
of a spin gap larger than 1~K in PCTO. Different \textit{T}-regimes can
still be distinguished. Upon cooling from high \textit{T}, 1/$T_{1}$
progressively decreases down to $\sim 20$~K where it shows a marked kink and
a steeper variation which fits to a sublinear power law $T^{0.4}$. This
change below $T\sim \theta $, corresponding to the saturation of the local
susceptibility, has to be related to the emergence of short range
spin-correlations. This evolution can be compared to the one of the 2D spin
liquid Herbertsmithite $T^{0.7}$~\cite{AO} and to the theoretical prediction
of power law behaviors for critical spin liquids~\cite{Hermele,Sachdev}.
However, contrary to these latter cases, the evolution of $1/T_{1}$ changes
here again below about 2~K where it sharply increases. The increase of $%
1/T_{1}$ below 2~K evidences a slowing down of the spin dynamics consistent
with the $\mu $SR results.

The compound PbCuTe$_{2}$O$_{6}$ appears as one rare example of a 3D AFM
exhibiting a dynamical ground state, \emph{i.e.} with no on-site frozen
moments. This is all the more striking since first principles calculations
suggest a high connectivity of the magnetic lattice ($z=8$) resulting from
three different AFM interactions of comparable strengths \cite{3D,STR}.
Whether these interactions compete and eventually enhance the magnetic
frustration as for instance in kapellasite~\cite{BF,EK} or reduce the strong
geometric frustration of the hyperkagome lattice generated by the dominant (%
\textit{n}.\textit{n}.) interaction requires a detailed study of the \textit{%
S} = 1/2 Heisenberg model with all three interactions. Despite the
complexity of the magnetic model, it is instructive to compare our results
to the prototype material Na$_{4}$Ir$_{3}$O$_{8}$ for the 3D quantum
hyperkagome model and related theories. Let us remind that in the case of
the iridate hyperkagome, spin-orbit coupling leads to an effective\textit{\ }%
$J_{\mathrm{eff}}$ = 1/2 \ Heisenberg model still under discussion, while
such a model is the natural starting point in PbCuTe$_{2}$O$_{6}$. Also, the
recent experimental work has shown that Na$_{4}$Ir$_{3}$O$_{8}$ experiences
a magnetic transition below $T=7$~K~\cite{MJG,ACS}, at variance with PbCuTe$%
_{2}$O$_{6}$ where no freezing has been detected. Now in the spin liquid
phases of both compounds, the NMR shifts are found to saturate at a rather
high value at low \textit{T}, a feature that is best accounted for in a
fermionic description of the magnetic excitations on the hyperkagome lattice
leading to a spinon Fermi surface and a constant Pauli-like susceptibility~%
\cite{MJLb,YZhou}. The absence of a large spin-gap in PbCuTe$_{2}$O$_{6}$
rules out the alternative possibilities of a valence bond crystal or a
topological spin liquid ground state suggested for the hyperkagome model~%
\cite{RM,MJLa}. In the fermionic framework, the $T^{\alpha }$ ($\alpha \sim
2 $ in strong applied field) behavior of the heat capacity observed below $%
\sim 0.6$~K~\cite{3D} is not predicted and requires an instability -namely a
partial gap opening- of the spinon Fermi surface. The strong slowing down of
the spin dynamics shown by the $\mu $SR and NMR results at about 1~K could
be the signature of such a crossover between two different spin liquids.
Further, a partial gapping of the spinon Fermi surface at $\sim 1$~K
resulting in a reduced density of spinon excitations may help in
understanding the weak field dependence of the gaussian-like $\mu $SR
relaxation, tentatively attributed to sporadic fluctuations below about 1~K.

To conclude, our investigations at low temperatures by magnetization, $\mu $%
SR, and NMR reveal that PbCuTe$_{2}$O$_{6}$ is a promising 3D
antiferromagnet with $S$ = 1/2\ where strong frustration leads to a spin
liquid behavior. This rare case invokes for an in depth investigation of the
appropriate magnetic Hamiltonian, including for instance high temperature
series expansion, together with theoretical developments in a fermionic
approach. In this context our results together with those in Ref.~\cite{3D}
give strong constraints on the possible ground states. In view of the
relatively weak coupling strength, the effect of substitutions, application
of external pressure, and local probe experiments at higher magnetic fields
might offer an appealing possibility to tune the magnetism of PCTO and to
explore further insights into its magnetic properties.

We acknowledge R. R. P. Singh for discussions on HTSE and P. Wiecki for some 
$T_{2}$ measurements. PK acknowledges support from the European Commission
through Marie Curie International Incoming Fellowship (PIIF-GA-2013-627322).
The research was supported by the U.S. Department of Energy, Office of Basic
Energy Sciences, Division of Materials Sciences and Engineering. Ames
Laboratory is operated for the U.S. Department of Energy by Iowa State
University under Contract No. DE-AC02-07CH11358. BK thanks DST INSPIRE
faculty scheme to carry out the research work. This work was also supported
by the French Agence Nationale de la Recherche under Grants
\textquotedblleft SPINLIQ\textquotedblright\ No. ANR-12-BS04- 0021, by
Universit\'{e} Paris-Sud Grant MRM PMP and by a SESAME grant from R\'{e}gion
Ile-de-France.

*khuntia@lps.u-psud.fr

\bigskip 

\textbf{Figure Captions}

Fig 1(Color online) (a) Temperature dependence of the magnetic
susceptibility in several applied fields. The bottom left inset shows a
partial view of the crystal structure of PbCuTe$_{2}$O$_{6}$. The top right
inset shows the \textit{T}-dependence of magnetic specific heat by \textit{T}
(\textit{C}$_{m}$/\textit{T}) in 0 T [adapted from Ref.\cite{3D}).

Fig. 2 (Color online) (a) Field dependence of muon polarization \textit{P}$%
_{z}$(\textit{t}) at $T=0.1$~K. The solid line correspond to dynamical
Kubo-Toyabe model (DKT) as explained in the text. (b) \textit{T}-dependence
of the relaxation rate obtained from the stretched exponential fit. The
inset shows the \textit{T}-dependence of the stretched exponent.

Fig. 3 (Color online) (a) Temperature evolution of field swept $^{207}$Pb
NMR spectra at 63.5 MHz (b) The temperature dependence of NMR signal
intensity and the inset shows the \textit{T} dependence of 1/\textit{T}$_{2}$
at 63.5 MHz.

Fig. 4 (Color online) (a) Temperature dependence of the$^{207}$Pb shift \
and the magnetic spin susceptibility $\chi _{spin}$. The solid lines
correspond to HTSE up to order 15 and 16 for the hyperkagome lattice and
extrapolations using Pad\'{e} approximation (adapted from~\cite{HTSE}). (b) 
\textit{T}- dependence of the spin-lattice relaxation rate at three
frequencies. The solid line is a fit to \textit{T}$^{\alpha }$ giving $%
\alpha =0.4\pm 0.05$.

\begin{center}
\bigskip 

\bigskip \textbf{SUPPLEMENTAL}
\end{center}

\subsection{\textbf{Spin Liquid State in the 3D Frustrated Antiferromagnet
PbCuTe}$_{2}$\textbf{O}$_{6}$\textbf{: NMR and }$\protect\mu $\textbf{SR
Studies}}

\subsubsection{\textbf{Magnetic Susceptibility}}

Magnetization measurements in different applied magnetic fields were
performed using a Quantum Design (QD) MPMS at high temperature. The
temperature dependence of magnetic susceptibility in the temperature range
0.45 K to 10 K measured in a commercial MPMS-7T SQUID magnetometer equiped
with a 
${{}^3}$%
He low temperature insert in various applied magnetic fields up to 7 T.

\subsubsection{\textbf{Muon Spin Relaxation (}$\protect\mu $\textbf{SR)}}

We have performed $\mu $SR measurements in zero field and in applied
magnetic fields at the Paul Scherrer Institute facility in a helium flow
cryostat on the GPS spectrometer and in the dilution fridge of the LTF
spectrometer.

Shown in Fig. S1 is the zero field muon $\mu ^{+}$ polarizations at
different temperatures down to 20 mK and the solid lines are fits to the
stretched exponential \textit{P}$_{z}$(\textit{t}) = \textit{P}$_{0}$ exp[-($%
\lambda $\textit{t})$^{\beta }]$ of the muon polarization. The relaxation
rate, $\lambda ,$\ thus obtained increases with decreasing temperature
suggesting the slowing down of spin fluctuations which is shown in Fig. 2 of
the main text of the manuscript. The constant value of $\lambda $\ beolw 0.5
K suggests the presence of a short-range spin correlations and dynamical
ground state inherent to the slow magnetic fluctuation of Cu$^{2+}$ spins.
The temperature dependence of stretched exponent ($\beta $) is shown as an
inset in Fig. 2(b) of the main text.

\subsubsection{\textbf{Nuclear Magnetic Resonance (NMR)}}

$^{207}$Pb (\textit{I} = 1/2; \ $\frac{\gamma _{n}}{2\pi }$=8.874 MHz/T) NMR
measurements were carried out using a home made phase-coherent spin-echo
pulse spectrometer in an extended temperature range\textit{\ }with a $^{3}$%
He-$^{4}$He dilution refrigerator (Kelvinox MX100, Oxford instruments) and
using a He flow cryostat installed at Ames Laboratory. The $^{207}$Pb-NMR
spectra were obtained by sweeping the magnetic field. As a further measure
to understand the low temperature magnetism, we have performed nuclear
spin-spin relaxation time ($T_{2}$) measurements (shown in the inset of Fig.
3(b) of the main text). The \textit{T}$_{2}$ value at each temperature was
obtained by fitting the transverse magnetization, $M_{xy}$\textit{\ }($t$) = 
$M_{0}$e$^{-2t/T_{2}}$, where $M_{0}$ is the initial magnetization and $t$\
is the delay between two \textit{rf} pulses ($\pi /2-t-\pi $). The raw NMR
signal intensity at\ each temperature was determined by extrapolating $%
M_{xy} $ to \textit{t} = 0 in the recovery of the transverse magnetization.
The signal intensity thus obtained was multiplied by temperature to
compensate for the Boltzmann factor. The temperature dependence of
normalized signal intensity in PbCuTe$_{2}$O$_{6}$ at 63.5 MHz \ thus
obtained is shown in Fig. 3(b) of main text of the manuscript. It is found
that 1/$T_{2}$ increases continuously on lowering the temperature, which is
contrary to conventional long range ordered antiferromagnet wherein one
expects a peak in 1/$T_{2}$.

\bigskip Shown in the left panel of Fig. S2 is the temperature dependence of
the NMR spin susceptibility compared with that of the bulk magnetic
susceptibility in the whole tempearure range of the present investigation.
The little increase of \textit{K}$_{spin}$ below 1 K is the shift of the
residual signal ascribed to a tiny fraction of defect spins present in the
polycrystalline sample.

The spin lattice relaxation time (\textit{T}$_{1}$) at each temperature was
determined by fitting the recovery of the nuclear magnetization \textit{M}$%
_{z}$(\textit{t}) = \textit{M} ($\infty $)(1--e$^{-t/T_{1}}$) towards its
equilibrium value \textit{M}($\infty $) after a single saturation pulse
(here \textit{M}$_{z}$(\textit{t}) and \textit{M}($\infty $) are the nuclear
magnetization at time \textit{t} after the saturation and equilibrium
nuclear magnetization at \textit{t}$\rightarrow \infty $ respectively). The
recovery of longitudinal magnetization $M_{z}$(\textit{t}) could
consistently be fitted with a single exponential function in the temperature
and magnetic field ranges of the present investigation, which indicates the
presence of single Pb site in the host lattice without structural
distortion. The NMR shift (\textit{K}) probes intrinsic spin susceptibility $%
\chi $($\mathit{q}$\textit{\ }= 0, $\omega $ = 0) of the system under
study.The spin-lattice relaxation rate tracks the low\ energy spin
fluctuations and is expressed as the mean of the imaginary part of dynamic
spin susceptibility $\chi ^{\prime \prime }(q,\omega )$ in the \textit{q}%
-space \textit{i.e}., 1/$T_{1}T\varpropto \sum \mid \mathit{A}_{hf}(q)\mid
^{2}\chi ^{\prime \prime }(q,\omega )$, where $A_{hf}$ ($q$) is the form
factor of the hyperfine interactions. The right panel of Fig. S2 represents
the temperature evolution of 1/\textit{T}$_{1}$ in the whole temperature
rage of the present investigation. As discussed in the main text, the peak
in 1/\textit{T}$_{1}$ at 1 K should not be taken as evidence of a magnetic
transition because it is concomitant with a huge loss of the signal
intensity. The decrease of 1/\textit{T}$_{1}$ below 1 K is rather associated
with the increasing of relative NMR signal from the tiny fraction of defect
spins which feature a longer \textit{T}$_{1}$.

\subsubsection{\textbf{Spin Susceptibility (Hyperkagome model)}}

The exchange interaction between the nearest neighbor Cu$^{2+}$ spins can be
expressed with Heisenberg Hamiltonian with $J$ as exchange coupling.

$H$ = $J$ $\sum_{<i,j>}S_{i}$.$S_{j}$

The strength of the exchange interaction ($J$) between the nearest neighbor
Cu$^{2+}$ spins on the hyperkagome lattice, was obtained from the fitting of
the magnetic susceptibility data following the high temperature series
expansion (HTSE) for $S$ = 1/2 system valid for hyperkagome lattice\cite{RRP}

$\chi $($T$) = $\frac{N_{A}g^{2}\mu _{B}^{2}}{4k_{B}}\sum_{0}^{16}b_{n}(-%
\frac{J}{4k_{B}T})^{n}$

with series expansion coefficients $b_{n}$ and the values of which can be
found in Ref.\cite{RRP}. The magnetic susceptibility and the intrinsic local
spin susceptibilty \ are well explained by the nearest-neighbor Heisenberg
model on a hyperkagome lattice. The HTSE extrapolations for magnetic
susceptibility using (7,7) and (8,7) Pad\'{e} approximants converge down to 
\textit{J}/4 \cite{RRP}. The fit of the magnetic susceptibility data to
hyperkagome model yields an AFM coupling $J$/$k_{B}$ = (14 $\pm $ 1) K
between Cu$^{2+}$ spins.

\textbf{Figure Captions}

SM Fig. 1 (Color online) The zero field muon spin polarizations at various
temperatures and the solid lines are fits to stretched exponential.

SM Fig 2. (Color online) Temperature dependence of the NMR spin
susceptibility ($K$) and the magnetic susceptibility ($\chi $) in the whole
temperature range of the present investigation. (b) \textit{T}-dependence of
the spin lattice relaxation rate in the whole temperature range of the
present work and the solid line is the \textit{T}$^{0.4}$ behavior.


\begin{thebibliography}{99}
\bibitem{FN} It is noticeable that even at very low $T$, $M$/$H$ is little
field dependent suggesting sizeable AFM interactions among all spins in the
sample and a small amount of quasi-free or "orphan spins" commonly observed
in frustrated materials.
\end{thebibliography}
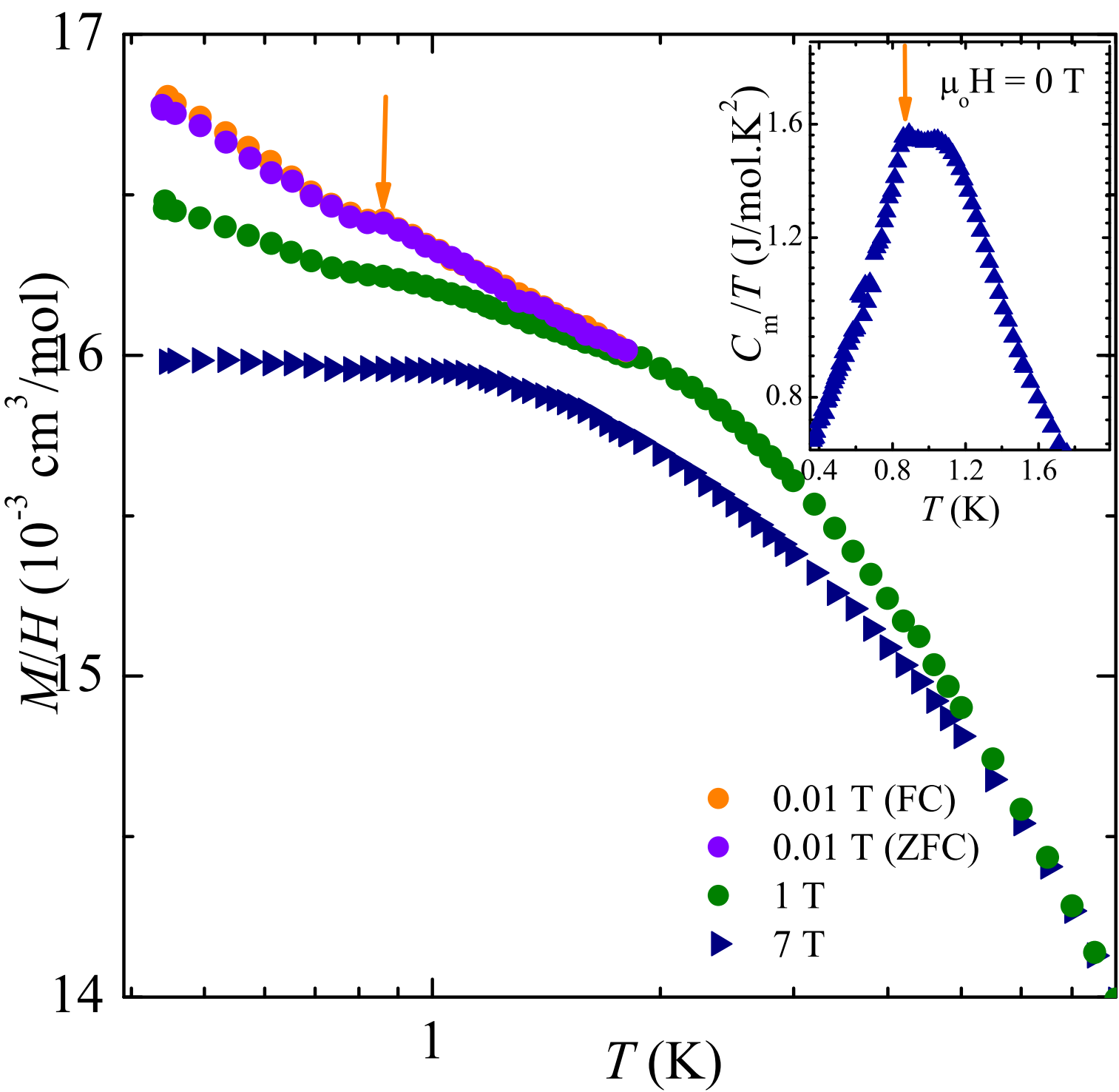
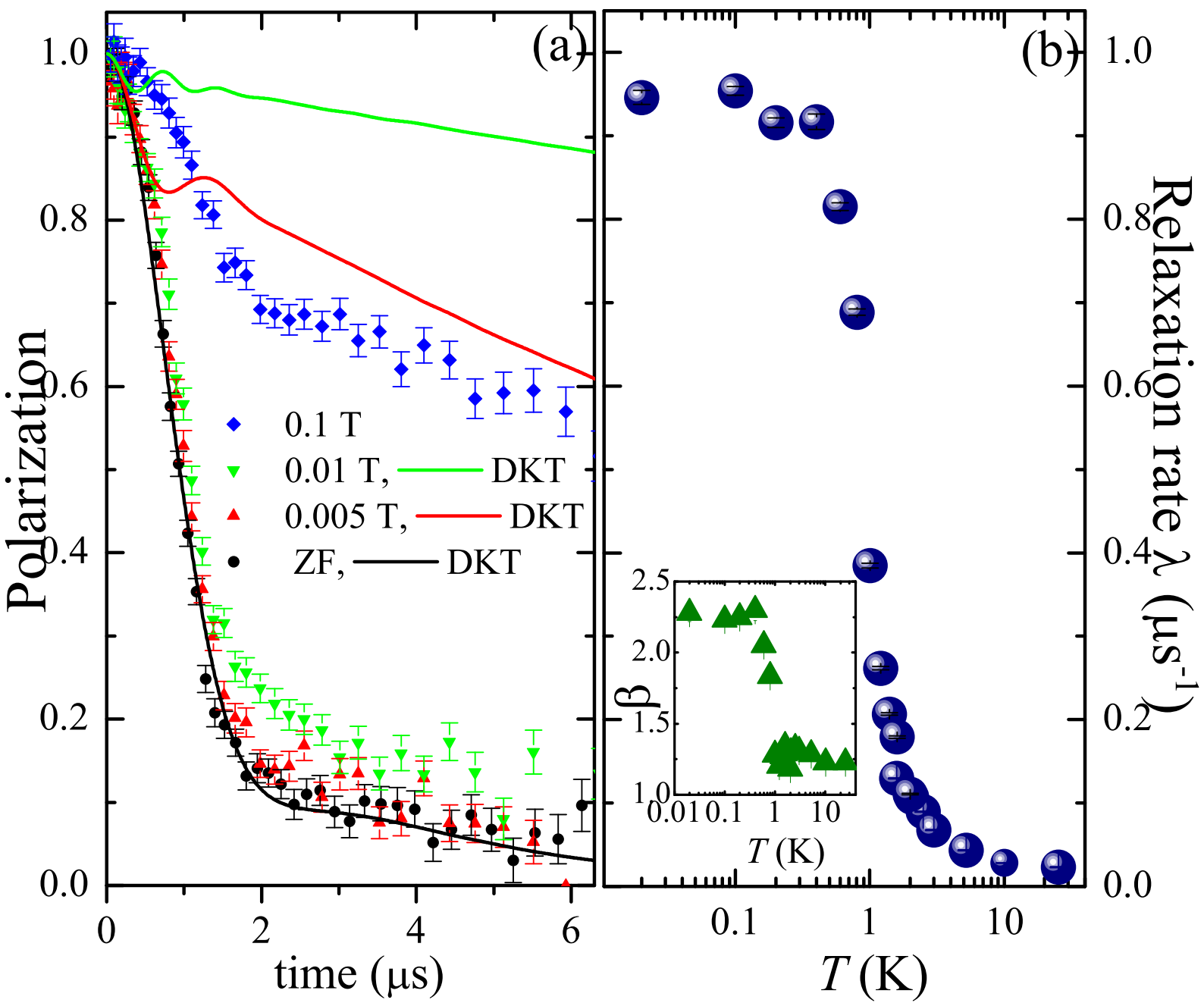
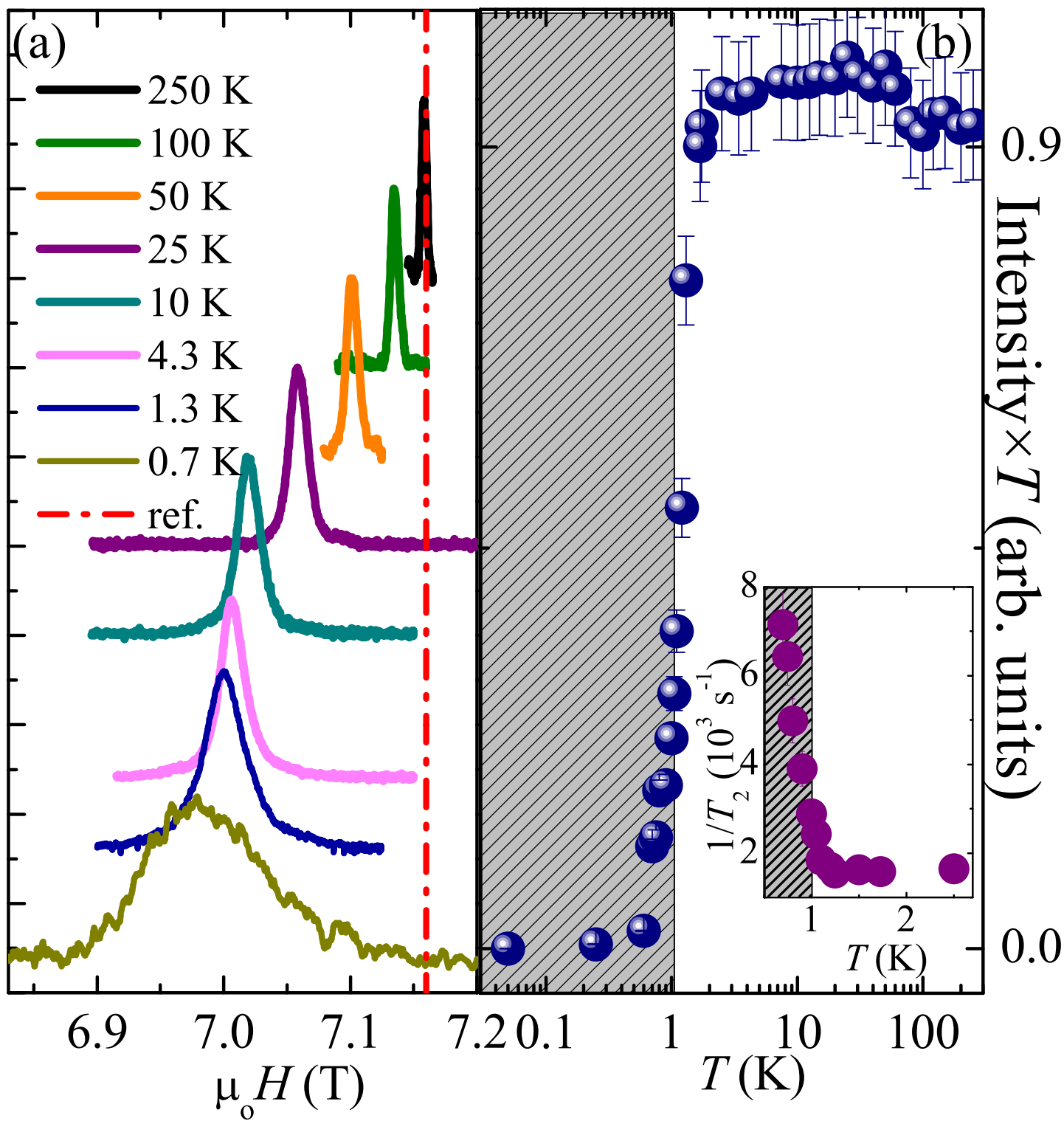
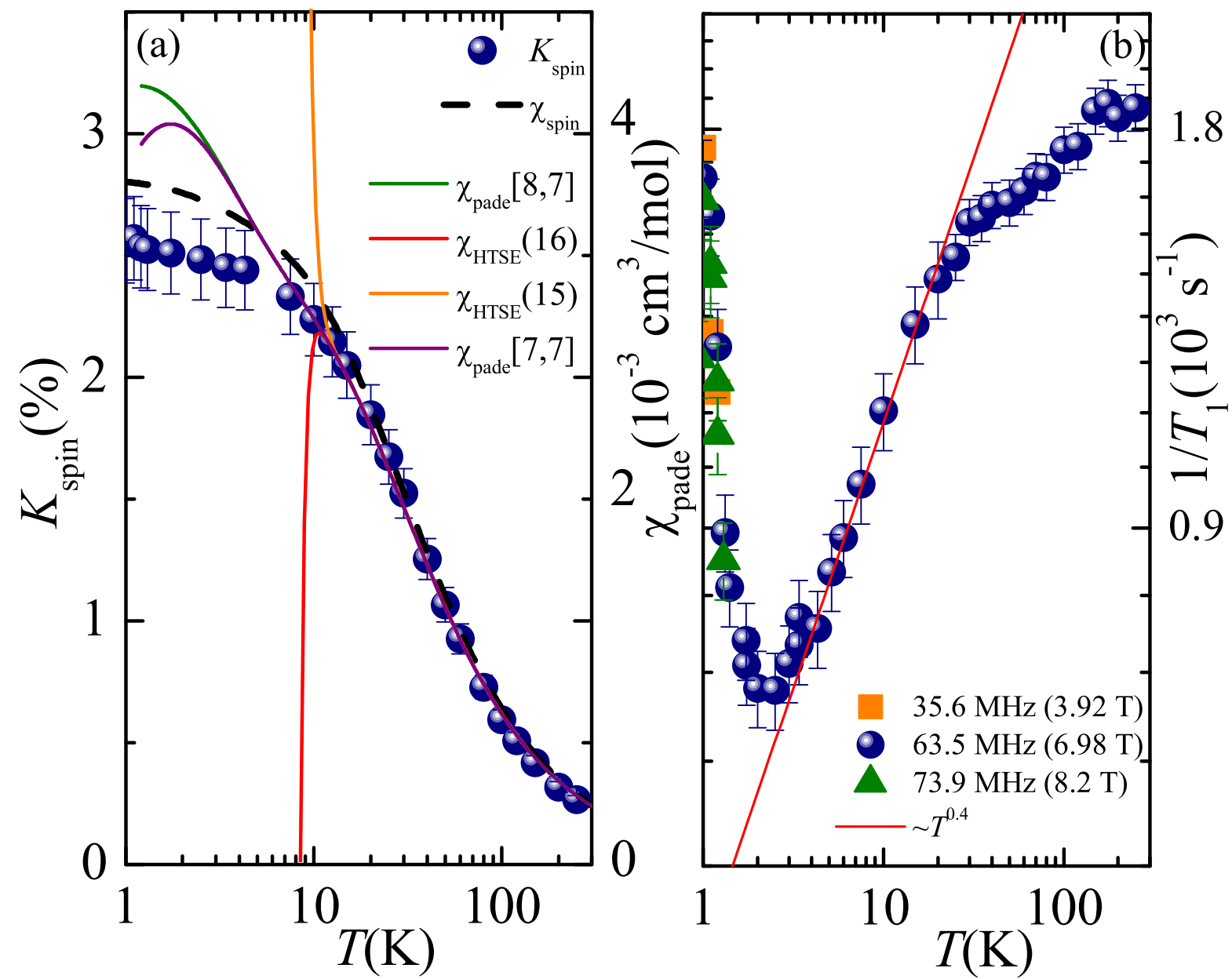
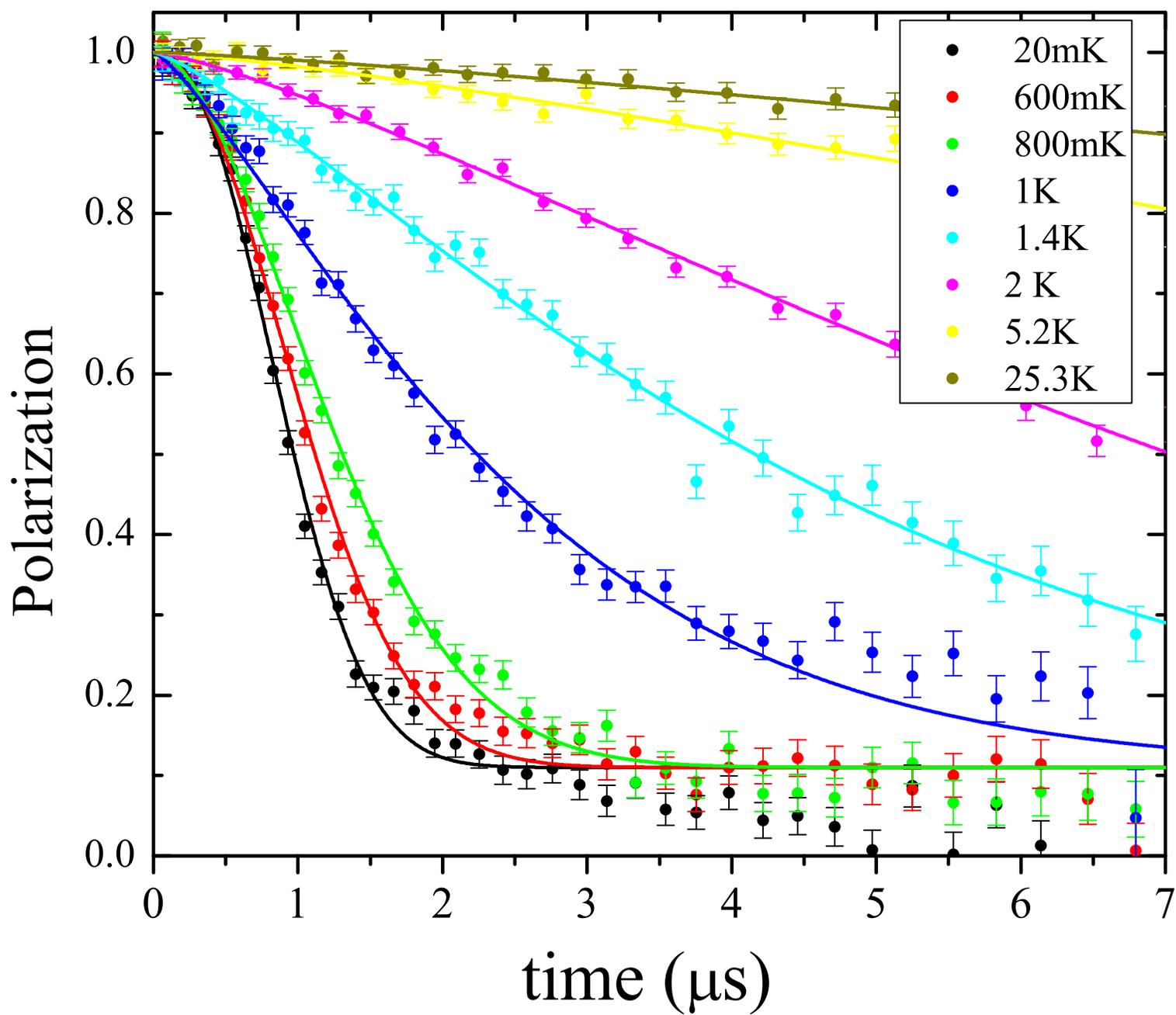
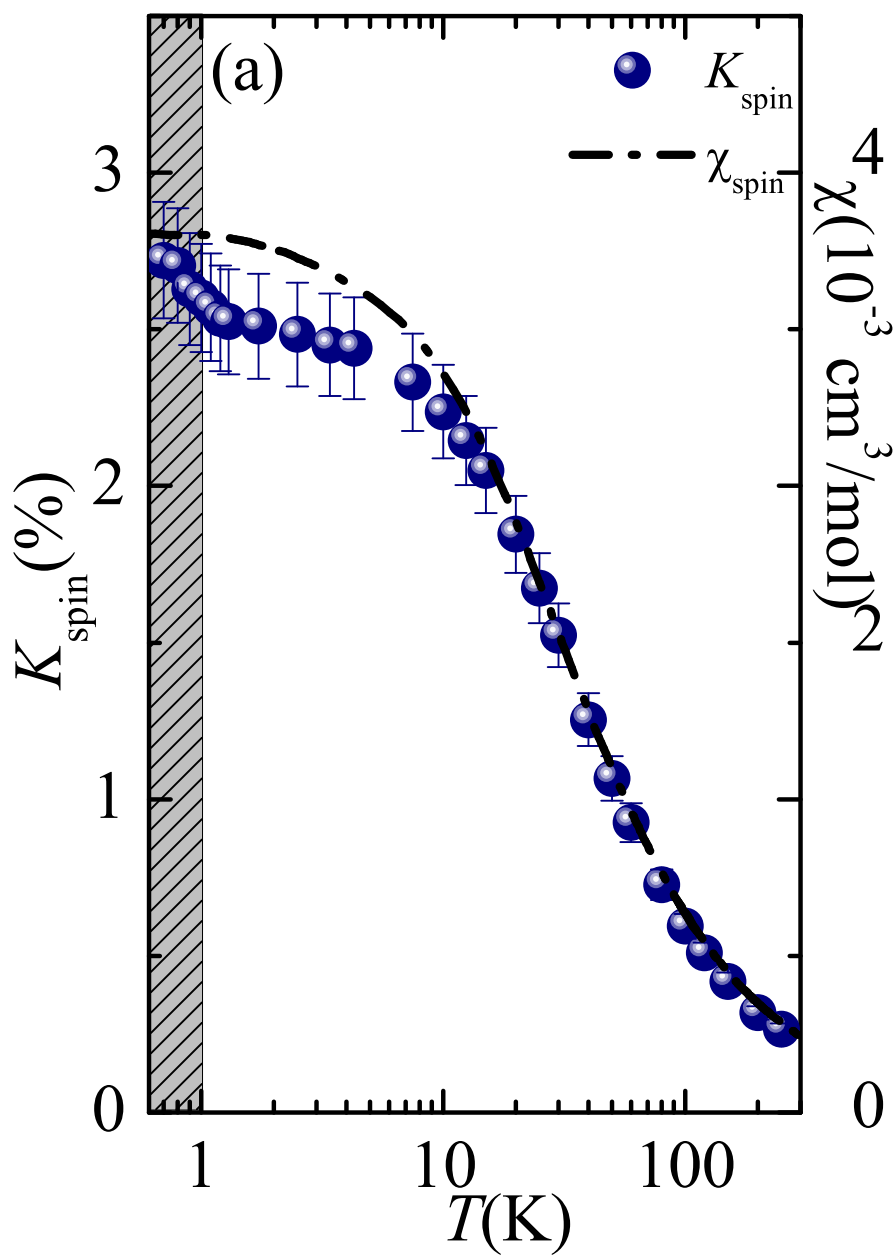
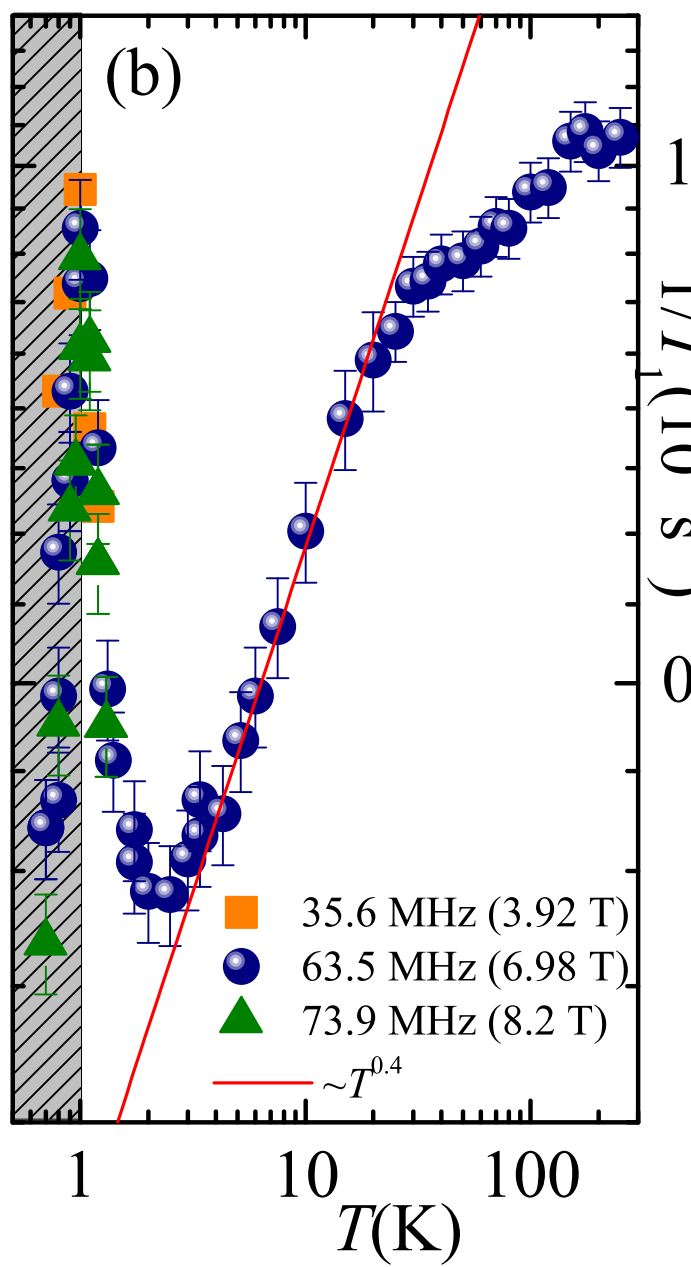
\end{document}